\documentclass[a4paper,12pt]{article}
\usepackage{amsmath}
\usepackage{amssymb}
\newcommand{\diag}{\operatorname{diag}}
\newcommand{\tr}{\operatorname{tr}}
\newcommand{\ch}{\operatorname{ch}}
\newcommand{\cs}{\operatorname{cs}}

\newcommand{\id}{{1}}
\newcommand{\Z}{{\mathbf{Z}}}
\newcommand{\C}{{\mathbf{C}}}
\newcommand{\real}{{\mathbf{R}}}

\newcommand{\calO}{{\mathcal{O}}}

\newcommand{\dd}{{\mathrm{d}}}
\newcommand{\ii}{{\mathrm{i}}}
\newcommand{\ee}{{\mathrm{e}}}
\newcommand{\U}{{\mathrm{U}}}

\newcommand{\GL}{{\mathrm{GL}}}
\newcommand{\step}{\vspace{.5em}}

\renewcommand{\L}{{\bar\psi}}
\newcommand{\R}{{\psi}}
\newcommand{\gammaL}{\Gamma_{\L}}
\newcommand{\gammaR}{\Gamma_{\R}}
\newcommand{\gammaLR}{\Gamma_{\R/\L}}
\newcommand{\Sigmai}{\Sigma^{(i)}}
\newcommand{\Treg}{{T^{2l}_{\mathrm{r}}}}

\newcommand{\wpow}[1]{^{#1}}

\newcommand{\PoL}{\Pi_{\L}}
\newcommand{\PoR}{\Pi_{\R}}
\newcommand{\PoLi}{{\PoL^{(i)}}}
\newcommand{\PoRi}{{\PoR^{(i)}}}
 
\def\s0#1#2{\mbox{\small{$ \frac{#1}{#2} $}}}
\def\eq#1{(\ref{eq:#1})}
\def\Eq#1{Eq.~(\ref{eq:#1})}

\begin{document}
\begin{center}

\thispagestyle{empty}

{\normalsize\begin{flushright}
DIAS-STP-02-02\\
FAU-TP3-02-16\\
hep-lat/0205005\\[12ex] 
\end{flushright}}

\mbox{\large \bf Chiral fermions on the lattice} \\[2ex]

{\it Dedicated to the 
memory of L.~O'Raifeartaigh}\\[5ex]

{Oliver Jahn
\footnote{jahn@itp.phys.ethz.ch}
$\!\!{}^{,a,b}$
and 
Jan M.~Pawlowski 
\footnote{jmp@theorie3.physik.uni-erlangen.de}
$\!\!{}^{,c}$}
\\[4ex]
{${}^a${\it Dublin Institute for Advanced Studies\\
10 Burlington Road, Dublin 4, Ireland}\\[1ex]
${}^b${\it 
Institut f\"{u}r Theoretische Physik, ETH Z\"{u}rich\\
CH-8093 Z\"{u}rich, Switzerland \footnote{present address}}\\[1ex]
${}^c${\it 
Inst. f\"ur Theoret. Physik III, Universit\"at Erlangen,\\ 
Staudtstra\ss e 7, D-91054 Erlangen, Germany.}}
\\[10ex]
{\small \bf Abstract}\\[2ex]
\begin{minipage}{14cm}{\small 
    We discuss topological obstructions to putting chiral fermions on
    an even dimensional lattice. The setting includes Ginsparg-Wilson
    fermions, but is more general. We prove a theorem which relates
    the total chirality to the difference of generalised winding
    numbers of chiral projection operators. For an odd number of Weyl
    fermions this implies that particles and anti-particles live in
    topologically different spaces.}
\end{minipage}
\end{center}
\newpage \pagestyle{plain} \setcounter{page}{1}

\noindent

\section{Introduction}
The formulation of lattice theories with chiral fermions has been a
long-standing problem \cite{Hasenfratz:2001bz},\cite{Creutz:2000bs} 
which is closely related to
the fermion doubling problem. It has been proven early on,
that it is impossible to maintain chiral symmetry or define a chiral
gauge theory on the lattice, if some basic assumptions are met. This
is the celebrated Nielsen-Ninomiya (NN) no-go theorem
\cite{Karsten:1980wd}-\cite{Friedan:nk}. It states that
each left-handed fermion on a lattice must be accompanied by a
right-handed fermion with the same quantum numbers. Thus, a 
lattice theory for a single Weyl fermion seems to be ruled out, but
also a lattice realisation of chiral symmetry since the latter would
assign opposite charges to left- and right-handed particles. \step

In turn, the Ginsparg-Wilson (GW) relation describes how close one
can get to the na\"{\i}ve chiral symmetry in a lattice theory
\cite{Ginsparg:1981bj}.  The derivation of the GW relation also
highlights how the no-go theorem could be circumvented. The GW
relation is derived from a block spin transformation of a lattice
action which enjoys the na\"{\i}ve chiral symmetry. In this sense it
defines the `natural' quantum chiral symmetry on the lattice. For a
long time, investigations of lattice theories with GW fermions have
been hampered by the fact that no explicit example was known of a
Dirac operator (in an interacting theory) obeying the GW relation.
This gap was filled in \cite{Neuberger:1997fp} (and, less explicitly, in
\cite{Hasenfratz:1998ri}). It also turned out that the action of GW
fermions is invariant under a generalised chiral transformation of the
fields \cite{Luscher:1998pq},\cite{Luscher:1998du}.  \step

Let us now focus on how the no-go theorem is avoided: a chiral
symmetry based on the GW relation fails to meet the key assumption of
the NN theorem, namely that the symmetries act on the fermions $\psi$ and
$\bar\psi$ as if they were (Dirac-) conjugates of each other. While
this is the case in Minkowski space-time, $\psi$ and $\bar\psi$ have
to be treated as independent fields in the Euclidean space-time used
for the formulation of lattice theories. Consequently, 
$\psi$ and $\bar\psi$ can transform independently under symmetry 
transformations. \step

In the present contribution we study the properties of general chiral
transformations on the lattice for the case of free fermions. We
derive a statement about
generalised chiral projections including those which are related to GW
Dirac operators.  To that end we consider Dirac operators $D$
which allow the definition of local chiral projections: $P_{\R}$ and
$P_{\L}$ with $D P_\R = P_{\L} D$ and $P_{\R/\L}^2= P_{\R/\L}$. Apart
from this, some technical conditions have to be met in order to ensure
the vanishing of lattice artifacts in the continuum limit.  Then the
following statement can be proven:
\begin{eqnarray*}
\chi= n[P_\R] - n[P_\L]
\end{eqnarray*}
where $n[P_{\R/\L}]$ is a (generalised) winding number of $P_{\R/\L}$
defined in \eq{winding} and $\chi$ is the total chirality of all
fermion species emerging in the continuum limit. It is also shown that
the windings $n[P_{\R/\L}]$ are integers.  As a corollary this implies
a version of the no-go theorem: for total chirality $\pm 1$ it is
impossible to find chiral projections $P_{\R/\L}$ with $P_\L=1-P_\R$.
The theorem proven here applies to even dimensional Euclidean lattices;
in particular it covers the case of 4-dimensional Euclidean lattices
instead of the 3-dimensional spatial lattices considered in
\cite{Nielsen:1980rz,Nielsen:1981xu,Friedan:nk}.  
Reference \cite{Karsten:1981gd} deals with 4-dimensional lattices but
restricts the form of the action rather strongly and does not allow
for momentum dependent chiral projections.  The latter, in particular,
has important consequences. A first account of the present work was given in
\cite{O'Raifeartaigh:mw}. \step 

The paper is organised as follows. In the second section we state the
theorem and some corollaries. We also discuss the
necessity and implications of certain properties imposed on the Dirac
operator and the projections. In the third section the winding numbers
$n[P_{\R/\L}]$ are evaluated and the theorem is proven.  We close with
a brief discussion of our findings. Some technical details are
deferred to the appendices together with an example which highlights
the difference between 3-dimensional (Hamiltonian) and 4-dimensional
(Euclidean) lattices.

\section{Theorem}\label{sec:theorem}
We state the theorem in Sect.~\ref{subsec:theorem} and discuss its
implications in Sect.~\ref{subsec:implications}.
\subsection{Setting and theorem}\label{subsec:theorem}
We consider free, massless fermions on an infinite $d$-dimensional
hyper-cubic lattice $\Lambda$ with lattice spacing $a$,
$\Lambda=\{n_{\mu}a|n_\mu\in\Z\}$, where $d=2 l$ is even. 
At each lattice site, two $M$-component spinors
$\psi(x),\bar\psi(x)\in\C^M$ are defined.  The action is given in terms
of a translationally invariant Dirac operator $D(x-y)\in\C^{M\times M}$,
\begin{equation}
  \label{eq:D}
  S = \sum_{x,y\in\Lambda} \bar\psi(x) D(x-y) \psi(y) \;.
\end{equation}
The Fourier transform of $D$,
\begin{equation}
  \label{eq:fourier}
  D(k) \equiv \sum_{x\in\Lambda} \ee^{-\ii k\cdot x} D(x) \;,
\end{equation}
is periodic with periods $2\pi/a$ on $\real^{2 l}$ and can therefore be
considered as a function from the $2l$-torus $T^{2 l}$ to $\C^{M\times
  M}$. The spinors $\psi,\bar\psi$ live in the spaces defined by the
constraints
\begin{equation}
  \label{eq:chiral-projection}
 P_\R(k) \psi(k) = \psi(k)
 \qquad\text{and}\qquad
  \bar\psi(k) P_\L(k) = \bar\psi(k)
\end{equation}
with translationally invariant Hermitean projection operators
$P_{\R/\L}=P_{\R/\L}^2$. This includes the trivial case
$P_{\L}=P_\R=\id$. For a theory with chiral fermions, the projections
$P_{\R/\L}$ are necessarily non-trivial, as one has to remove part of
the degrees of freedom.  We would also like to emphasise that $P_\R$
and $P_\L$ can be different since in a Euclidean theory $\psi$ and
$\bar\psi$ are independent fields. We will prove the following:\\[2em]
{\bf Theorem:} Given a lattice theory in $2l$ dimensions with the
action \eq{D} and projections $P_\L$, $P_\R$ as in
\eq{chiral-projection}, where the Dirac operator $D$ and the
projection operators $P_{\R/\L}$ have the properties (i)-(iii):
\begin{itemize}
\item[(i)] \emph{locality:}\quad $|D_{i j}(x)|,|P_{\L\,i
    j}(x)|,|P_{\R\,i j}(x)|<c\,\ee^{-|x|/\lambda}$ for some real
  constants $c,\lambda$.  This implies that $D(k)$ and $P_{\R/\L}(k)$
  are analytic in $k_\mu$ in a strip around the real axis. 
\item[(ii)] {\it spin-$\frac12$ zeros:}\quad the real poles $k^{(i)}$
  of the propagator $D(k)^{-1}$ have the form
  \begin{subequations}
    \label{eq:poles}
    \begin{gather}
      \label{eq:pole-D}
      D(k)^{-1} =
      \frac{(k_\mu-k^{(i)}_{\mu})}{|k-k^{(i)}|^2} \, {\Sigma^{(i)}_{\mu}}^\dagger +
      \text{finite} \;, \\
      \label{eq:S-alg}
      \begin{aligned}
        {\Sigma^{(i)}_\mu}^\dagger \Sigma^{(i)}_\nu
        + {\Sigma^{(i)}_\nu}^\dagger \Sigma^{(i)}_\mu
        &= 2 \delta_{\mu\nu} \, \PoRi
        &\quad\text{with}\quad \PoRi &= \PoRi^2 = \PoRi^\dagger \;,
        \\
        \Sigma^{(i)}_\mu {\Sigma^{(i)}_\nu}^\dagger
        + \Sigma^{(i)}_\nu {\Sigma^{(i)}_\mu}^\dagger
        &= 2 \delta_{\mu\nu} \, \PoLi
        &\quad\text{with}\quad \PoLi &= \PoLi^2 = \PoLi^\dagger \;, 
      \end{aligned}
    \end{gather}
  \end{subequations}
\item[(iii)] \emph{compatibility of chiral projections:}\quad the
  chiral projection operators and the Dirac operator satisfy
  \begin{equation}\label{eq:chiral}
    D\, P_{\R} = P_\L\, D \;.
  \end{equation}
\end{itemize}
Then the total chirality of all fermion species 
in the continuum limit is given
by
\begin{equation}
  \label{eq:theorem}
  \chi = n[P_\R] - n[P_\L] \;.
\end{equation}
where
\begin{equation}\label{eq:winding}
 n[P] \equiv \frac{1}{l !}\left(\frac{i}{2\pi}\right)^{l} 
\int_{T^{2 l}} \tr\,P (\dd P)^{2l} 
\in \Z \;. 
\end{equation} 
Moreover $n[P]$ is a topological invariant with integer values for local,
translationally invariant, Hermitean projection operators $P$ on the Fourier 
space $T^{2l}$. $\square$\vspace{.3cm}

\noindent The theorem implies that
\begin{enumerate}
\item a non-zero total chirality is only possible with non-trivial
  projections ($P_\R\ne1$ or $P_\L\ne1$ or both), 
\item an odd total chirality is only possible if the spaces onto which
  $P_\L$ and $P_\R$ project are neither identical nor orthogonal,
  i.e., $P_\L\ne P_\R$ and $P_\L\ne1-P_\R$, because otherwise
  $n[P_\L]=n[P_\R]$ or $n[P_\L]=-n[P_\R]$.  The spaces onto which they
  project (and their orthogonal complements) are in fact inequivalent
  fibre bundles over $T^{2 l}$.
\label{it:asym}
\end{enumerate}
The theorem does not exclude even non-zero chirality
$\chi=2n[P_\R]$ if $P_\L=1-P_\R$.  This is in contrast to the
situation for 3-dimensional spatial lattices, where only zero
chirality is possible \cite{Nielsen:1980rz}-\cite{Friedan:nk}.  An
example with $\chi=4$ in 2 dimensions (and $\chi=16$ in 4 dimensions)
presented in App.\ \ref{app:example} shows, that this is realised.\step 
 
GW fermions \cite{Ginsparg:1981bj} are included in \eq{chiral} as
they are defined with
\begin{eqnarray}\label{eq:GW} 
\ \  \{\gamma_{2l+1},\,D\}= a D\,\gamma_{2l+1}\,D,\quad {\rm where} \quad 
\gamma_{2l+1}= \ii^l \gamma_1\cdots \gamma_{2l}\quad {\rm with}
 \quad \gamma_{2l+1}^2=\id.
\end{eqnarray}
Thus, they admit the (non-unique) definition of $P_{\R/\L}$ with
$P_\R=\s012 (\id -\gamma_{2l+1})$ and $P_\L=\s012 (\id
+\gamma_{2l+1}(1-a D))$. One can prove with \eq{GW} that $P_{\R/\L}$
satisfy \eq{chiral}.  Moreover $P_\R^2=P_\R$ and $P_\L^2=P_\L$. A 
similar analysis applies to Dirac operators satisfying a recently
discussed generalisation of the GW relation \cite{Fujikawa:2000my}.

\subsection{Necessity and implications of the properties (i)-(iii)}
\label{subsec:implications}
Before proving the theorem in the next section, we first would like to 
elaborate a bit on the properties (i)-(iii): 

\bigskip

Locality (i) of $D$ and the chiral projections guarantees that the
continuum limit of the lattice theory does not depend on the details
of the discretisation.  In a local theory, only the behaviour at the 
zeros of $D$ matters in the continuum limit.

\bigskip

The structure of the zeros (ii) is
determined by the requirement that the fields in the continuum limit
carry spin-$\frac12$ representations of the Euclidean group.  A pole
of $D^{-1}$ of the form \eq{poles} gives rise to a continuum action density
$\bar\psi\,k\cdot\Sigma\,\psi$ (we suppress the superscript ${(i)}$ on
$\Sigma$).
$\PoR$ and $\PoL$ project onto the right and left eigenspaces with
vanishing eigenvalues of $D$, i.e.\ onto those components of $\psi$
and $\bar\psi$ that survive the continuum limit. Like the
$\gamma$-matrices, the $\Sigma_\mu$ define spin-$\frac12$
representations of the rotation group $SO(2l)$. The fermions
$\bar\psi,\psi$ live in the two different representations generated by
\begin{equation}
  \label{eq:Sigma}
  \begin{aligned}
    \tfrac12 \Sigma^{\R}_{\mu\nu} 
    &\equiv \tfrac14 ( \Sigma_\mu^\dagger \Sigma_\nu
    - \Sigma_\nu^\dagger \Sigma_\mu ) \;,\\
    \tfrac12 \Sigma^{\L}_{\mu\nu} 
    &\equiv \tfrac14 ( \Sigma_\mu \Sigma_\nu^\dagger
    - \Sigma_\nu \Sigma_\mu^\dagger ) \;. 
  \end{aligned}
\end{equation}
The matrix $\Sigma_\mu$ couples the representations $\Sigma^\L$ and
$\Sigma^\R$ to a vector.  Therefore, the continuum action
$\bar\psi\,k\cdot\Sigma\,\psi$ is rotationally invariant.
In turn, fermions in the spin representations \eq{Sigma} and the
requirement of the correct continuum limit lead to a pole structure of
$D^{-1}$ as in \eq{poles}.\step 

It follows from \eq{S-alg} that we can write $\Sigma$ more explicitly
as
\begin{eqnarray}\label{eq:S}
  \Sigma^{(i)}_{\mu} = U^{(i)} \,
  \diag[ \underbrace{\sigma_\mu,\dots,\sigma_\mu}_{n_+}\,,\,
  \overbrace{\sigma_\mu^\dagger,\dots,\sigma_\mu^\dagger}^{n_-} ] 
  \, {V^{(i)}}^\dagger,  
\end{eqnarray}
where the $\sigma_\mu$ and $\sigma_\mu^\dagger$ form right- and
left-handed $2^{l-1}$-dimensional irreducible representations of
(\ref{eq:S-alg}) without projections.  These are unique up to
bi-unitary transformations. Here, right-handed means
$\ii^l\sigma_1^\dagger\sigma_2\cdots\sigma_{2l-1}^\dagger\sigma_{2l}=+1$,
which implies that $\sigma^\dagger$ is left-handed,
$\ii^l\sigma_1\sigma_2^\dagger\cdots\sigma_{2l-1}\sigma_{2l}^\dagger=-1$.
For $2l=4$,
one can choose $\sigma_4=1$ and $\sigma_i=\ii\tau_i$ where $\tau_i$
are the Pauli matrices. We also have
\begin{eqnarray}\label{eq:PLR}
  \PoLi =U^{(i)} {U^{(i)}}^\dagger,\! \quad\! 
  \PoRi =V^{(i)} {V^{(i)}}^\dagger\quad {\rm with}  
  \quad {V^{(i)}}^\dagger V^{(i)}={U^{(i)}}^\dagger U^{(i)}
  = \id_{2^{l-1}(n_++n_-)}. 
\end{eqnarray}
Thus, a pole of the form (\ref{eq:poles}) gives rise to $n_+$ right-
and $n_-$ left-handed fermions in the continuum limit, where $n_+$ and
$n_-$ are the number of $\sigma_\mu$ and $\sigma_\mu^\dagger$ in
$\Sigma^{(i)}_\mu$ respectively.  The corresponding components of
$\psi$ and $\bar\psi$ respectively are obtained as eigenspaces for
eigenvalues $\pm1$ of the chirality operators
\begin{equation}
  \label{eq:gamma-5-i}
  \gammaR^{(i)} \equiv 
  \ii^l \, {\Sigma^{(i)}_{1}}^\dagger \Sigma^{(i)}_{2}
  \cdots {\Sigma^{(i)}_{2l-1}}^{\!\!\!\!\!\dagger\,\,\,} \Sigma^{(i)}_{2l} \;,
  \quad \quad  \gammaL^{(i)} \equiv 
  \ii^l \, {\Sigma^{(i)}_{1}} {\Sigma^{(i)}_{2}}^\dagger
  \cdots {\Sigma^{(i)}_{2l-1}} {\Sigma^{(i)}_{2l}}^\dagger \;.
\end{equation}
The $\gammaLR^{(i)}$ are Hermitean and have eigenvalues $\pm1$ and $0$. 
Taking into account the projection \eq{chiral-projection}, the total
chirality of all fermion species in the continuum limit is given
by
\begin{equation}
  \label{eq:chirality}
  \chi
  = \frac{1}{2^{l-1}} \sum_i \tr\bigl[ P_\R(k^{(i)})\,\gammaR^{(i)} \bigr] \;.
\end{equation}

The possibility of vector-like (Dirac) zeros is contained in (ii):
if the $\Sigma_\mu$ are Hermitean, \eq{S-alg} turns into the standard
anti-commutation relations for $\gamma$-matrices in the image of
$\PoR=\PoL$.

\bigskip

The compatibility condition (iii) for $D$ and the chiral projections
is, for local $P_{\R/\L}$, equivalent to the following compatibility
condition for $P_\R$ and the spin representations $\Sigma^{\R\,(i)}$ at
the zeros $k^{(i)}$,
\begin{equation}
  \label{eq:compat-Sigma-P}
  \bigl[ \Sigma^{\R\,(i)}_{\mu\nu}, P_\R(k^{(i)}) \bigr] = 0 \;;
\end{equation}
the form of $P_\L$ then follows from that of $P_\R$ via \eq{chiral}.
Hence the property \eq{chiral} is, at its root, only a constraint at
the points $k^{(i)}$. It is for this reason, that the proof of the
theorem boils down to the calculation of windings at these points.
The relation \eq{compat-Sigma-P} ensures that the projected spinor
$P_\R\psi$ also transforms under $SO(4)$; it contains complete
irreducible components of the representation $\Sigma^{\R\,(i)}$ only. It
also follows that
\begin{equation}\label{eq:Pcom}
  \bigl[\PoR,\, P_\R(k^{(i)})\bigr] = 0.
\end{equation}
To prove that \eq{chiral} implies \eq{compat-Sigma-P} and \eq{Pcom},
first note that it implies $P_\R D^{-1}=D^{-1}P_\L$.  Since $P_{\R/\L}$
are analytic, we can expand in powers of $q\equiv k-k^{(i)}$ to get
\begin{equation}
  \label{eq:pole-comm}
  P_\R(k^{(i)}) \, \Sigma_\nu^\dagger = \Sigma_\nu^\dagger \, P_\L(k^{(i)})
\end{equation}
for all $\nu$, where we have used that we are free to choose $q_\mu =
\delta_{\mu\nu}|q |$ (we have suppressed the superscript $(i)$ on $\Sigma$
again). Using \eq{pole-comm} and its 
Hermitean conjugate one concludes that $\Sigma_\mu^\dagger
\Sigma_\nu P_\R(k^{(i)})=P_\R(k^{(i)})\Sigma_\mu^\dagger \Sigma_\nu$. 
With \eq{S-alg} and \eq{Sigma}
this leads to \eq{compat-Sigma-P} and \eq{Pcom}. \step 

Conversely, for any local $P_\R$ satisfying \Eq{compat-Sigma-P}, we
can define $P_\L\equiv D P_\R D^{-1}$. Except for the zeros of $D$,
analyticity of $P_\L$ follows in the set where $P_\R$ and $D$ have
this property. Equations \eq{compat-Sigma-P} and \eq{S-alg} imply
$P_\R(k^{(i)}) \Sigma_\mu^\dagger = P_\R(k^{(i)}) \Sigma_\mu^\dagger
\Sigma_\nu \Sigma_\nu^\dagger = \Sigma_\mu^\dagger \Sigma_\nu
P_\R(k^{(i)}) \Sigma_\nu^\dagger$ for $\nu\ne\mu$ (no sum).  As
$D\Sigma_\mu=\calO(k)$, the poles of $D^{-1}$ drop out and $P_\L$ is
finite and analytic there as well.  It goes without saying that
similar statements and relations like
\eq{compat-Sigma-P}, \eq{Pcom}, \eq{pole-comm} follow for
$\Sigma^\L_{\mu\nu}$, $\PoL$, $P_\L(k^{(i)})$.\step 

Local projection operators $P_{\R/\L}$ satisfying \eq{chiral} are also 
relevant in non-chiral theories where the constraints
\eq{chiral-projection} are not needed: they can be used to define
charges $Q_{\R}\equiv1-2P_\R$, $Q_{\L}\equiv2P_\L-1$ and a `chiral'
symmetry
\begin{equation}
  \label{eq:chiral-sym}
  \begin{aligned}
    \psi &\to \ee^{\ii\alpha Q_{\R}} \psi \;, \\
    \bar\psi &\to \bar\psi \, \ee^{\ii\alpha Q_{\L}}  \;.
  \end{aligned}
\end{equation}
The existence of such a symmetry with local charges, however, implies
the existence of local projections only if their eigenvalues are
non-zero and non-degenerate in the entire Brillouin zone, for instance
if the charges are integer-valued.  The theorem presented in this
paper applies to such a symmetry as well.  It implies that a symmetry
of this kind that goes over to the standard chiral symmetry in the
continuum limit necessarily acts on $\psi$ and $\bar\psi$ in an
asymmetric way (not as if $\psi$ and $\bar\psi$ were Dirac conjugates
of each other).  Note, however, that symmetries with charges whose
eigenvalues vanish somewhere can be useful in non-chiral theories,
e.g.\ \cite{Luscher:1998pq}, \cite{Fujikawa:2000my}.  The theorem does
not make a statement about these symmetries.  Indeed, the symmetries 
used in \cite{Luscher:1998pq,Fujikawa:2000my}, essentially, 
lead to projections that
satisfy $P_\L=1-P_\R$ as well as \eq{chiral} although the image of
$P_\R$ contains only a single left-handed fermion.  These projections
are discontinuous at some points in momentum space, so they are not
local. \step

The chirality of a fermion in the continuum limit is
determined by $\gammaLR^{(i)}$ in \eq{gamma-5-i}
 rather than $P_\R$ or $P_\L$. The
images of the latter may well contain fermions of different chirality.
Hence, our setting includes theories with any number of left- and
right-handed fermions (in particular vector-like theories for which we
can set $P_\R=P_\L=1$).  If only one chirality is desired, one has to
require that $\gammaLR^{(i)}$ has a definite sign in the image of
$P_\R(k^{(i)})$.  This is not necessary for our theorem, so we do not
make this requirement. However, we still use the term `chiral' 
projections for $P_{\R/\L}$ since they treat left- and right-handed
fermions differently in general.\step 

Accordingly, also the charges $Q_{\R/\L}$ need not coincide with the
chirality operator $\gammaLR^{(i)}$ at the zeros of $D$. Equation
\eq{compat-Sigma-P} only guarantees that $Q_\R(k^{(i)})$ and
$\gammaR^{(i)}$ are simultaneously diagonalisable, as are
$Q_\L(k^{(i)})$ and $\gammaL^{(i)}$. Equation \eq{chiral-sym} goes over
to the standard chiral symmetry in the continuum limit only if
$\gammaLR^{(i)}$ has only the eigenvalues 1 and 0 in the image of
$P_{\R/\L}(k^{(i)})$.

\bigskip

We close the section with an explicit -- and relevant -- example, the
overlap Dirac operator \cite{Neuberger:1997fp} in four dimensions. It is a GW
Dirac operator, see \eq{GW}, and, for vanishing gauge field, it is
given by
\begin{equation}
  \label{eq:neuberger}
  a D = 1 - \cos\theta + \ii \gamma \cdot \hat p \,
  \sin\theta 
\end{equation}
where $\hat p=p/|p|$, $p_\mu(k)$ and $\theta(k)$ are periodic functions
in momentum space and $\gamma_\mu$ a (fixed) set of Dirac matrices.
Local chiral projections can be defined as $P_\L =
\tfrac12(1+\gamma_5)$ and $P_\R=\tfrac12(1-Q_{\R})$ where
\cite{Luscher:1998du}
\begin{equation}
  \label{eq:gamma-luescher}
  Q_{\R} = \gamma_5(1-a D) = 
 \gamma_5 \cos\theta - \ii\gamma_5 \gamma \cdot \hat p \,
  \sin\theta \;.
\end{equation}
Regarding $-\ii\gamma_5\gamma_\mu$ and $\gamma_5$ as basis vectors in a
5-dimensional space, $Q_{\R}$ takes values on the unit sphere in this
space.  The form of Eq.~(\ref{eq:winding}) suggests that $n[P_\R]$
measures the degree, or winding number, of the map $Q_{\R}\colon T^4\to
S^4$.  The degree can be expressed as the
number of times a fixed point on the target is taken, weighted by the
orientation (the sign of the Jacobian), provided the Jacobian does not
vanish at the chosen point.  We may choose the point $Q_{\R}=\gamma_5$,
i.e., $\theta=0$.  This corresponds to the zeros of $D$.
If $D$ has only a single zero, $Q_{\R}$ has unit winding number.  For
the overlap operator this can be explicitly verified by studying the
functions $\theta$ and $p_\mu$.  So we find $n[P_\R]=-1$ and $n[P_\L]=0$
and verify the theorem~\eq{theorem}.  For general functions $\theta(k)$
and $p_\mu(k)$, the orientation of a zero is given by (minus) the
winding number of $\hat p\colon S^3\to S^3$ around the zero.  Since
this coincides with the definition of chirality in
(\ref{eq:chirality}), the theorem holds true.

\section{Proof of the theorem}
\label{sec:proof}
First, in Sect.~\ref{sec:integer}, we prove that the 
winding number $n[P]$ is an integer. Then, in Sect.~\ref{sec:proof}, 
we show that $n[P_\R]-n[P_\L]$ in 
\eq{theorem} can be transformed into the formula for the total chirality 
as given on the rhs of \eq{chirality}.

\subsection{The winding number $n[P]$}
\label{sec:integer}

The power of the theorem depends crucially on the fact that $n[P]$ is
an integer for local projection operators $P(k)$. Hence, before tackling
the proof of the relation~\eq{theorem} we argue that $n[P]\in\Z$. This
discussion will also shed some light on the interpretation of the
invariant $n[P]$. To that end we express $P$
in terms of an orthonormal basis
$\Psi=(\psi_1,...,\psi_N)\in\C^{M\times N}$ (where $N=\tr P$ is the
rank of $P$),
\begin{equation}
  \label{eq:basis}
  P \equiv \Psi \Psi^\dagger
  \qquad\text{with}\qquad
  \Psi^\dagger \Psi = \id_N
  \;.
\end{equation}
Then, a general basis is given by $\Psi^v\equiv\Psi\, v$ with
$v\in\U(N)$.  Since $P$ is periodic in $k_\mu$, $\Psi$ in general
satisfies the boundary conditions
\begin{equation}
  \label{eq:bc}
  \Psi(k+\hat a_\mu) = \Psi(k) \, u_\mu(k)
  \qquad\text{with}\qquad
  u_\mu(k) \in \U(N) \quad {\rm and} \quad 
  (\hat a_\mu)_\nu=\s0{2\pi}{a} \delta_{\mu\nu}\;.
\end{equation}
The transition functions $u_\mu$ defined in this way satisfy the
cocycle conditions  
\begin{equation}
  \label{eq:cocycle}
  u_\mu(k) \, u_\nu(k+\hat a_\mu) 
  = u_\nu(k) \, u_\mu(k+\hat a_\nu). 
\end{equation}
They define a $\U(N)$ fibre bundle over $T^{2l}$. 
If this fibre bundle is non-trivial, the eigenspace of $P$ does not
admit a globally smooth basis.  We will see that $n[P]$ measures (part
of) this non-triviality.

To characterise the fibre bundle, we define the $\U(N)$ gauge
potential and field strength
\begin{equation}
  \label{eq:A}
  A = \Psi^\dagger \, \dd \Psi \;,
  \qquad
  F = \dd A + A \wedge A \;.
\end{equation}
They obey the expected boundary conditions
\begin{align}
  \label{eq:A-bc}
  A(k+\hat a_\mu)
  &= u_\mu^\dagger(k) \, \bigl( A(k) + \dd \bigr) \, u_\mu(k) \;, \\
  F(k+\hat a_\mu) &= u_\mu^\dagger(k) \, F(k) \, u_\mu(k) \;. 
\end{align}
It follows from (\ref{eq:basis}) and (\ref{eq:A}) that, for even
dimension $d=2 l$, the winding number $n[P]$ as defined in
\eq{winding} is given by the integral of the $l$th Chern character
(cf.~App.~\ref{sec:chern}):
\begin{equation}
  \label{eq:chern-gamma}
  n[P] = \int \ch_{l} (F)\;.
\end{equation}
In general, this is not an integer but only a multiple of
${1}/{l!}$.  For $\U(N)$ bundles over $T^{2l}$, however, it is an
integer.  This follows directly from the Atiyah--Singer index theorem.
There, the $l$th Chern character for a torus $\U(N)$-bundle is shown to
be identical to the index of the related Dirac operator, which is an
integer.\step 

In the light of the discussion above there is a natural interpretation
of (\ref{eq:chiral}) as a map between inequivalent $\U(N)$-bundles over
the torus $T^{2l}$. For highlighting this fact and for
later use in Sect.~\ref{subsec:proof} let us discuss this in more
detail. On $\Treg\equiv T^{2l}\setminus\{k^{(i)}\}$ we can write Eq.\ 
(\ref{eq:chiral}) as
\begin{equation}
  \label{eq:gammaLR-D}
  P_\L =  \varepsilon(D)\,P_\R\, \varepsilon^\dagger(D)
  \quad\textrm{with}\quad
  \varepsilon(D) \equiv D (D^\dagger D)^{-1/2} \;.
\end{equation}
$\varepsilon(D)$ is unitary. In \eq{gammaLR-D} we have used that
$[P_\R, (D^\dagger D)^{-1/2}]=0$, which follows directly from
\eq{chiral} and its Hermitean conjugate.  Two orthonormal bases
$\Psi_\L$ of $P_\L$ and $\Psi_\R$ of $P_\R$ are therefore related by
\begin{equation}
  \label{eq:Psi-L}
  \Psi_\L = \varepsilon(D) \Psi_\R g
  \quad\text{with}\quad
  g=\Psi_\R^\dagger\varepsilon^\dagger(D)\Psi_\L \in \U(N) \;.
\end{equation}
The function $g$ is continuous except at the zeros $k^{(i)}$ of $D$
where $\varepsilon(D)$ is ill-defined.  It is not periodic but
satisfies the boundary conditions
\begin{equation}
  \label{eq:g-bc}
  g(k+\hat a_\mu) = {u^{\R\,\dagger}_\mu}(k) \, g(k)\, u^\L_\mu(k) \;.
\end{equation}
where $u^\R_\mu$ and $u^\L_\mu$ are the transition functions of
$\Psi_{\R}$ and $\Psi_\L$.  These are thus related by
\begin{equation}\label{eq:relateus}
  u^\L_\mu(k) = g^\dagger(k) \, u^\R_\mu(k) \, g(k+\hat a_\mu) \;.
\end{equation}
Hence, if $g$ (restricted to the boundary) carries non-trivial
topology, the two sets of transition functions $u^{\R/\L}$ define
inequivalent $\U(N)$-bundles. However, since $g$ is smooth outside of
the zeros of $D$, its non-trivial content can be extracted from its
windings at $k^{(i)}$, which are introduced by $\varepsilon(D)$.

\subsection{Proof of equation \eq{theorem}}\label{subsec:proof}

The discussion in the previous section already suggests that the
difference $n[P_\L]-n[P_\R]$ is directly related to a homotopy class
of the map $\varepsilon(D)$ around the zeros of $D$. Indeed, we shall
see that it is given by
\begin{eqnarray}\label{eq:cherntowinding}
  n[P_\L]-n[P_\R]=\sum_i \nu_i[ g]
  \end{eqnarray}
where 
\begin{eqnarray}\label{eq:nu}
\nu_i[g]=\lim_{\epsilon\to 0} b_{2l-1} 
\int_{\|k-k^{(i)}\|=\epsilon } 
\tr\, (g^{-1} \dd g)^{2l-1}\quad {\rm with} 
\  b_{2l-1}=(-1)^{l-1} \frac{(l-1)!}{(2l-1)!} 
\left( \frac{i}{2\pi } \right)^{l}\;
\end{eqnarray}
is the $\pi_{2l-1}$ winding of $g\in\GL(N)$ about $k^{(i)}$.  The
gauge function $g$ is given in terms of $\varepsilon(D)$ by \eq{Psi-L}. \step 

From \eq{cherntowinding} only a few technical steps have to be invoked
in order to prove \eq{theorem}.  For the proof of \eq{cherntowinding}
we resort to the fact that the difference of winding numbers is given
by the integral of the difference of Chern character $\ch_{l}$, see
\eq{chern-gamma}:
\begin{equation}\label{eq:cherns}
  n[P_\L]-n[P_\R] = \int \ch_{l}[F_L]-\int \ch_{l}[F_R], 
\end{equation}
where $F_L,F_R$ are the field strengths of $A_\L=A[P_\L]$ and
$A_\R=A[P_\R]$ respectively.  The integral over a Chern character is a
topological invariant of the underlying fibre bundle.  It can
therefore be calculated with any gauge field obeying the same boundary
conditions, in particular we can replace $A_\L$ by $\tilde
A_\L=A_\R^g\equiv g^{-1}(A+\dd)g$.  Now we use that $\ch_l$ can
(locally) be related to its Chern-Simons form $\cs_{2l-1}$
by $\ch_{l}[F]=\dd \cs_{2l-1}[A,F]$.  Furthermore
\begin{eqnarray}\label{eq:alpha}
\cs_{2l-1}[A^g,F^g]-\cs_{2l-1}[A,F]=
\cs_{2l-1}[g^{-1}\dd g,0]+\dd\alpha_{2l-2}[A,F,\dd g\,g^{-1}] \;.
\end{eqnarray} 
Upon differentiation, the second term drops out. With 
$\cs_{2l-1}[A,0]=b_{2l-1}\tr A^{2l-1}$ (see e.g.\ 
\cite{Nakahara}, page 400) and integrating over the Brillouin zone
$I^{2l}\equiv[-\pi/a,\pi/a]^4$, we find
\begin{eqnarray}
  n[P_\L] - n[P_\R] = \nu[g],  
\end{eqnarray}
where
\begin{eqnarray} 
  \nu[g]=b_{2l-1} \int_{\partial I^{2l}} 
  \tr\, (g^{-1} \dd g)^{2l-1}
\end{eqnarray}
is the total $\pi_{2l-1}$ winding on $\partial I^{2l}$. Thus,
\eq{cherntowinding} follows with the remark, that $g$ is continuous
except for the zeros of $D$ and the density of $\nu$ is closed. Then,
$\nu[g]$ equals the sum of the winding numbers of $g$: 
$\nu[g] =\sum_i \nu_i[g]$.\step 

It is left to relate $\nu_i[g]$ to the chirality $\chi$.  First note
that in $g=\Psi_\R^\dagger\,\varepsilon^\dagger(D)\Psi_\L$ one can
replace $\Psi_\R$ and $\Psi_\L$ by their values at $k^{(i)}$ and
$\varepsilon^\dagger(D)$ by $D^{-1}$,
\begin{eqnarray}\label{eq:tildeg}
  \nu_i[g] = \nu_i[g_i]\quad {\rm  with}\quad 
  g_i=  
  \Psi_\R^\dagger(k^{(i)}) D^{-1} \Psi_\L(k^{(i)}) \;.
\end{eqnarray}
The first replacement is allowed because the bases $\Psi_{\R/\L}$ can
be chosen with finite derivatives (at least locally) and higher-order
terms drop out in \Eq{nu} in the limit $\varepsilon\to0$;  the second
because the set of positive definite Hermitean matrices is
contractible, so $\sqrt{D^\dagger D}$ can be deformed to $\id$. \step

Now we employ the explicit form of $D$ about its zeros $k^{(i)}$.
Using \eq{chiral} and the cyclicity of the trace we arrive at
\begin{eqnarray}\label{eq:arrive}
\tr\,(g_i^{-1}\,\dd g_i)^{2l-1}=
-\tr\,\left(P_\R(k^{(i)}) J\right)^{2l-1},\quad {\rm where}\quad 
J\equiv D^{-1} \,\dd D.
\end{eqnarray}
Intuitively one expects that only those parts of $D$ and $D^{-1}$ can
contribute to $\nu_i[g_i]$ that carry the Dirac structure
$q\cdot \Sigmai$ and $ |q|^{-2}\, q\cdot {\Sigmai}^\dagger$, respectively ($q\equiv
k-k^{(i)}$).  Indeed we find
\begin{eqnarray}\label{eq:usings}
  \nu_i[g_i]
  =-
  \lim_{\varepsilon\to 0}b_{2l-1}
  \int_{\|k-k^{(i)}\|=\varepsilon}\tr\,P_\R (k^{(i)})
  {J_0}^{2l-1}, 
\end{eqnarray}
where 
\begin{equation}
  \label{eq:J-chiral-zeros'}
  J_{0} \equiv \frac{1}{|q|^2} \, q \cdot {\Sigmai}^\dagger \,
  \dd q\cdot \Sigmai \;. 
\end{equation}
As the derivation of \eq{usings} is a bit technical we defer it to
Appendix~\ref{technical}. For performing the integration in \eq{usings}
we use the symmetry properties of the integral.  We infer from
\eq{J-chiral-zeros'} that
\begin{equation}
  \label{eq:J-cubed-chiral}
  J^{2l-1}_{0} = \frac{1}{|q|^{2l}} 
  \left( 
    (-1)^{l-1} q \cdot {\Sigmai}^{\dagger} \, \dd q \cdot \Sigmai \,
    \bigl( 
      \dd q \cdot {\Sigmai}^{\dagger} \, \dd q \cdot \Sigmai \bigr)^{l-1} 
    + \calO(q\cdot\dd q)
  \right) 
\end{equation}
where the algebra~(\ref{eq:S-alg}) has been used and $\calO(q\cdot\dd
q)$ denotes terms containing $q\cdot\dd q$ as a factor.  These do not
contribute when integrated over 3-spheres centred at $q=0$.  The
first term yields
\begin{equation}
  \label{eq:J-cubed-int}
   \lim_{\varepsilon\to0} b_{2l-1}\int_{|k-k^{(i)}|=\varepsilon}
  J_{0}^{2l-1}
  = \frac{\ii^l}{2^{l-1}}\frac{1}{(2l)!} \epsilon_{\mu_1\cdots\mu_{2l}} \,
  \Sigma_{\mu_1}^{(i)\dagger} {\Sigma_{\mu_2}^{(i)}}
  \cdots  \Sigma_{\mu_{2l-1}}^{(i)\dagger} \Sigma_{\mu_{2l}}^{(i)}
  = \frac{1}{2^{l-1}} \gammaR^{(i)} \;.
\end{equation}
By inserting (\ref{eq:J-cubed-int}) into (\ref{eq:usings}) and the
latter into \eq{tildeg}, we obtain
\begin{equation}
  \label{eq:n-q-chiral}
  n[P_\L] - n[P_\R] 
  = -\frac{1}{2^{l-1}} \sum_i \tr \bigl[ P_\R(k^{(i)}) \gammaR^{(i)} \bigr]
  = -\chi \;.
\end{equation}
Thus $n[P_\R]-n[P_\L]$ is given by the total chirality of all
fermion species appearing in the eigenspace of $P_\R$. $\blacksquare$ 

\bigskip 

For the proof we have employed some cohomology theory, many
readers might be unfamiliar with. It is possible to avoid the use of
\eq{alpha} at the expense of tedious calculations. The integrand of
the difference $n[P_\L]-n[P_\R]$ is a total derivative
$d\omega_{2l-1}$, which can be explicitly calculated by using
$P_\L=D^{-1} P_\R D$ on $\Treg=T^{2l}\setminus \{k^{(i)}\}$, following
from \eq{chiral}. Then $n[P_\L]-n[P_\R]=\int_{\partial
  \Treg}\omega_{2l-1}$.  In four dimensions ($l=2$) it
follows after a straightforward, but rather lengthy calculation, that
on $\Treg$, the integrand in $n[P_\L]-n[P_\R]$ can be written as
\begin{eqnarray}\label{eq:divergence}
  \dd \omega_3 = \tr D\, P_\R\, D^{-1} \bigl(\dd(D\,P_\R\, D^{-1} )\bigr)^4
  - \tr P_\R (\dd P_\R)^4 
\end{eqnarray}
with 
\begin{eqnarray}\label{eq:omega}
  \omega_3 \equiv - \tr \left[ J\,(J\,P_\R)^2-\tfrac23 (J\,P_\R)^3- 
    2 (J\,P_\R)^2 \dd P_\R +J\,P_\R\, J\, \dd P_\R-2 J\, P_\R (\dd P_\R)^2
  \right].
\end{eqnarray}
Then, one proceeds by exploiting the properties of $J$ on $\partial
\Treg=\bigcup_i \{k|\,\| k-k^{(i)}\|=\epsilon\}$ as discussed in
Appendix~\ref{technical}. Since $P_\R$ is smooth and $J$ has only
linear singularities, only the first two terms on the rhs of \eq{omega}
can contribute. On $\partial\Treg$ it follows that
$\omega_3|_{\partial\Treg}= -\s013\tr P_\R(k) J_0^3+\calO(q^{-2})$.
Then one proceeds as from \eq{usings}. We add that even
deriving \eq{omega} is quite tedious and it gets increasingly
complicated when going to higher dimensions.  Moreover, within this approach 
the underlying topological structure gets obscured.

\section{Conclusions}
We have investigated the topological obstructions to implementing 
chiral symmetry on Euclidean lattices in general even dimensions. Our
findings are summarised in section~\ref{sec:theorem} in terms of a
theorem. Its setting allows for general {\it local} chiral projections
and Dirac operators, which, as a specific case, includes
Ginsparg-Wilson fermions \cite{Ginsparg:1981bj}. Within this setting
the total chirality $\chi$ is given by the difference of winding
number $n[P_\R]$ and $n[P_\L]$:
\begin{eqnarray*} 
  \chi= n[P_\R] - n[P_\L] \;.
\end{eqnarray*}
Here, $P_{\R/\L}$ are the projection operators defining the spaces of
fermions: $P_\R \psi=\psi,\ \bar\psi P_\L=\bar\psi$, see
\eq{chiral-projection}. The invariants $n[P]$ were shown to be
integers. \step 
 
This constitutes a generalisation of the Nielsen-Ninomiya no-go
theorem.  Let us briefly recapitulate the setting and the consequences
of the original theorem and its variants. In
\cite{Karsten:1980wd,Nielsen:1980rz,Nielsen:1981xu,Friedan:nk}
Hamiltonian (spatial) lattices and general symmetric projections are
considered, whereas \cite{Karsten:1981gd} deals with Euclidean
lattices and constant projections. Then, in both settings, the no-go
theorem states that a non-vanishing total chirality $\chi$ is
excluded. \step

In turn, the present formulation of the theorem implies that the
projections $P_\L$ and $1-P_\R$ have to be topologically inequivalent
for odd chirality. Hence, in this case symmetric projections $P_\L\neq
1-P_\R$ cannot be used. In particular this rules out symmetric
projections for one Weyl fermion: $\chi=1$. \step

Evidently, the realisation of an even number of only left handed
fermions with symmetric projections $P_\L=1-P_\R$ is not excluded by
$\chi= n[P_\R] - n[P_\L]$. However, if such a theory could be realised, the
projections cannot be constant. For constant projections
both winding numbers $n[P_\R]$ and $n[P_\L]$ vanish and the total
chirality is zero, in accordance with the no-go theorem of  
\cite{Karsten:1981gd}. 
Consequently, this additional option, if at all,
can only be realised for momentum-dependent chiral projections. In
Appendix~\ref{app:example} we present an example for such a case. The
example indicates that, as in the case of doubling modes, 
$\chi=2^{2 l}$ is needed if one requires invariance under the discrete Euclidean
group. 

\section{Acknowledgements}
We thank H.~Banerjee, C.~Nash and V.\,V.~Sreedhar for discussions.
Most of this work was done in collaboration with L. O'Raifeartaigh,
who sadly passed away in November 2000. His enthusiasm, deep insight
and more generally his humanity are greatly missed.

JMP thanks the Dublin Institute for Advanced Studies for hospitality
and financial support. The work of OJ has been supported by the
Alexander von Humboldt Foundation through a Feodor Lynen Research
Fellowship (grant number V-3-FLF-1068701).

\appendix

\section{Example for $\chi\ne0$}\label{app:example}

In this appendix, we provide an example of a lattice theory with an
even number of fermions of the \emph{same} chirality and
$P_\L=1-P_\R$.  According to Ref.\ \cite{Friedan:nk}, this is not
possible on 3-dimensional, spatial lattices.  It is not ruled out by
\eq{theorem} for even dimensional space-time lattices, however, and the
example shows that it can indeed occur.  To keep expressions simple,
we actually present a 2-dimensional example and indicate how it can be
generalised to 4 dimensions.  Our theorem also holds in 2 dimensions,
where the winding number \eq{winding} takes the form 
\begin{equation}
  \label{eq:winding-2d}
  n[P] \equiv \frac{\ii}{2\pi} \int_{T^2} \tr \bigl[ P (\dd P)^2
  \bigr] \; 
\end{equation}
and the total chirality is given by
$\chi=\sum_i\tr\bigl[P_\R(k^{(i)})\gammaR^{(i)}\bigr]$ with
$\gammaR^{(i)}=\ii {\Sigma_1^{(i)}}^\dagger \Sigma_2^{(i)}$. 
Now consider the na\"{\i}ve Dirac operator
\begin{equation}
  \label{eq:D-naive}
  D = \gamma_\mu \sin k_\mu
\end{equation}
(we have set the lattice spacing $a$ to $1$) and define chiral
projections $P_\R\equiv\frac12(1-Q_{\R})$ and $P_\L\equiv
1-P_\R$ with
\begin{equation}
  \label{eq:gamma3R}
  Q_{\R} \equiv \frac {\gamma_3 \cos k_1 \cos k_2
    + \gamma_1 \sin k_2 - \gamma_2 \sin k_1}
  {\sqrt{1 + \sin^2 k_1 \sin^2 k_2}} 
\end{equation}
with $\gamma_3\equiv\ii\gamma_1\gamma_2$.  It easy to see that
$Q_{\R}^2=1$ and $Q_{\R}^\dagger=Q_{\R}$, so that $P_{\R/\L}$ are
projections, and that $D Q_{\R}=-Q_{\R}D$, so that \Eq{chiral} holds
with $P_\L=1-P_\R$.  Furthermore $Q_{\R}$ is analytic in a cylinder
around the plane of real $k_i$, so $P_{\R/\L}$ are as well.

Near the zeros of $D$, $k^{(i)}_\mu=i_\mu\pi$ with the multi-index
$i_\mu=0,1$,
\begin{equation}
  \label{eq:D-zero}
  D = (-1)^{i_1} \gamma_1 q_1 + (-1)^{i_2} \gamma_2 q_2 + \calO(q^2)
\end{equation}
where $q_\mu=k_\mu-k_\mu^{(i)}$.
We read off $\Sigma_\mu=(-1)^{i_\mu}\gamma_\mu$ and find
\begin{equation}
  \label{eq:gamma3i}
  \gammaR^{(i)} = (-1)^{i_1+i_2} \gamma_3 \;.
\end{equation}
On the other hand, \Eq{gamma3R} gives
\begin{equation}
  \label{eq:gamma3Ri}
  Q_{\R}(k^{(i)}) = (-1)^{i_1+i_2} \gamma_3 \;.
\end{equation}
So $P_\R$ projects onto the left-handed component of $\psi$ at
all zeros of $D$.  The total chirality after chiral projection is
$\chi=-4$, the theory contains 4 species of left-handed fermions.\step

Equation \eq{gamma3R} can be generalised to 4 dimensions as follows:
we put $\cos\theta=\cos k_1 \cos k_2 \cos k_3 \cos k_4$, so that
$\sin\theta$ still cancels the poles at $\sin k_\mu=0$.  The fraction
is the scalar product of a unit vector $t_\mu$ with $\gamma_\mu$.  In
order for $Q_{\R}$ to anticommute with $D$, $t_\mu$ has to be
orthogonal to the vector $p_\mu=\sin k_\mu$.  This can be achieved
with the choice $t=(p_2,-p_1,p_4,-p_3)/|p|$.  The resulting projected
theory contains 16 species of right-handed fermions.  Note that this
construction is not possible in 3 dimensions: since $t$ depends only
on $p/|p|$ and is orthogonal to $p$, it can be considered as a vector
field of unit length on $S^3$; on $S^2$, however, all vector fields
vanish somewhere, so such a $t$ can not exist.  This argument does not
exclude a different construction, of course.

The above example does not contradict Ref.~\cite{Karsten:1981gd}. There, 
the chiral charge $Q_\R$ is assumed to be momentum independent.

\section{Derivation of \eq{usings}}\label{technical}

Here we present the technical details of the derivation of \eq{usings}. 
In view of \eq{nu} and \eq{tildeg}, \Eq{usings} is equivalent to 
\begin{eqnarray}\label{eq:usingsapp}
\lim_{\varepsilon\to 0}\int_{\|k-k^{(i)}\|=\varepsilon}\tr\,\left(
P_\R (k^{(i)}){J}\right)^{2l-1}=
\lim_{\varepsilon\to 0}\int_{\|k-k^{(i)}\|=\varepsilon}\tr\,P_\R (k^{(i)})
J_{0}^{2l-1}, 
\end{eqnarray}
where $J_{0}=|q|^{-2} \, q \cdot \Sigma^\dagger \, \dd q\cdot \Sigma$ 
as defined in 
\eq{J-chiral-zeros'}.  We have dropped the index $^{(i)}$ since we
focus on one zero of $D$.  Only the singular pieces of $J$ can 
contribute to the integral on the lhs of \eq{usingsapp}. 
For tracking them down we write $D$ and $D^{-1}$ 
in an expansion about $k^{(i)}$ as given in \eq{pole-D}: 
\begin{eqnarray}\label{eq:represent}
D(k) =
      q_\mu \Sigma_\mu+ \tilde M(k), \quad \quad 
D(k)^{-1} =
      \frac{q_\mu}{|q|^2} \Sigma_\mu^\dagger + M(k), 
\end{eqnarray}
where we have put $q\equiv k-k^{(i)}$ and the matrices
$M$ and $\tilde M$ have analytic
entries. We are only interested in the divergent term in
$J=\frac{1}{|q|^2}q\cdot \Sigma^\dagger\,\dd D+\calO(|q|^0)$. With  
\eq{S-alg} it follows that $\Sigma^\dagger= \PoR\Sigma^\dagger \PoL$. 
Moreover 
the projection operators $\Pi_{\R/\L}$ commute with $P_\R(k^{(i)})$
(see \eq{Pcom}). Hence for calculating the lhs of \eq{usingsapp} only the 
singular piece of $J\,\PoR$ is required. 
Consequently, we are only interested 
in the non-vanishing part of $\PoL\,\dd D \, \PoR$.  To obtain it,
consider
\begin{equation}
  q\cdot \Sigma = q \cdot \Sigma D^{-1} D \PoR 
  = \PoL D \PoR + q\cdot \Sigma (\tilde M \PoR + q\cdot \Sigma)
\end{equation}
and
\begin{equation}
  q\cdot \Sigma = D D^{-1} q\cdot \Sigma 
  = \tilde M \PoR + \calO(q) \;.
\end{equation}
These imply $\PoL D \PoR=q\cdot \Sigma+\calO(q^2)$, and we find
\begin{equation}
  \label{eq:J-chiral-zerosapp}
  J\, \PoR =   J_{0} + \calO(|q|^0)
  \quad\text{with}\quad
  J_{0} 
  \equiv \frac{1}{|q|^2} \, q \cdot \Sigma^\dagger \, \dd q\cdot \Sigma \;.
\end{equation}
Furthermore it follows from \eq{pole-comm} that $[J_0,\,P_\R(k^{(i)})]=0$. 
Thus, \eq{usingsapp} follows.

\section{Chern characters in terms of $P$}
\label{sec:chern}

We show that the winding number $n[P]$ is given by the integrated
Chern character of the fibre bundle associated with $P$,
see \Eq{chern-gamma}.  To this end, we use
\begin{equation}
  F = \dd \Psi^\dagger \wedge \dd \Psi
  + \Psi^\dagger \dd \Psi \wedge \Psi^\dagger \dd \Psi
\end{equation}
and
\begin{equation}
  (\dd P)^2 \Psi 
  = (\dd \Psi \, \Psi^\dagger + \Psi \,\dd\Psi^\dagger )^2 \Psi 
  = \Psi \, \dd \Psi^\dagger \wedge (1 - \Psi \Psi^\dagger ) \,\dd\Psi 
  = \Psi F    
\end{equation}
to find
\begin{equation}
  \label{eq:chern-P}
  \tr \bigl[ F\wpow{l} \bigr] 
  = \tr \bigl[ \Psi^\dagger \Psi F\wpow{l} \bigr]
  = \tr \bigl[ \Psi^\dagger (\dd P)\wpow{2l} \Psi \bigr]
  = \tr \bigl[ P (\dd P)\wpow{2l} \bigr]\;.
\end{equation}
The Chern characters can now be expressed as
\begin{equation}
  \ch_l(F) 
  \equiv \frac{1}{l!} \tr \biggl[ \left(\frac{\ii F}{2\pi} \right)^{\!l\,} \biggr]
  = \frac{1}{l!} \left( \frac{i}{2\pi} \right)^{\!l} \tr \bigl[ P (\dd
  P)\wpow{2l} \bigr]
  \;,
\end{equation}
which coincides with the integrand in the definition \eq{winding} of
$n[P]$.

\end{document}